\documentclass[lettersize,journal]{IEEEtran}
\usepackage{amsmath,amsfonts}
\usepackage{algorithmicx}
\usepackage{algorithm}
\usepackage{array}
\usepackage[caption=false,font=normalsize,labelfont=sf,textfont=sf]{subfig}
\usepackage{textcomp}
\usepackage{stfloats}
\usepackage{url}
\usepackage{verbatim}
\usepackage{graphicx}
\usepackage{cite}

\usepackage{hyperref} 
\usepackage{booktabs} 
\usepackage{multirow} 
\usepackage{makecell}
\usepackage{algpseudocode}

\begin{document}

\title{SS4Rec: Continuous-Time Sequential Recommendation with State Space Models}

% \author{Wei Xiao\textsuperscript{1,3,*}, Huiying Wang\textsuperscript{1,2,*}, Qifeng Zhou\textsuperscript{1,2,†}, and Qing Wang\textsuperscript{4}

\author{Wei Xiao\textsuperscript{*}, Huiying Wang\textsuperscript{*}, Qifeng Zhou\textsuperscript{†}, and Qing Wang

\thanks{Manuscript created XXXX, XXXX. This work was supported in part by the National Natural Science Foundation of China (Grants No. 62171391). % We sincerely appreciate Dr. Qing Wang’s helpful revisions, which significantly enhanced the quality of this work.
\textit{(Corresponding author: Qifeng Zhou.)}}

\thanks{Wei Xiao is with the Department of Automation, Xiamen University, Xiamen 361005, China (email: xiaowei2002103@foxmail.com).}
\thanks{Huiying Wang and Qifeng Zhou are with the Department of Automation, Xiamen University, Xiamen 361005, China, and also with the Xiamen Key Laboratory of Big Data Intelligent Analysis and Decision-making, Xiamen University, Xiamen, China (e-mail: huiyingwang@stu.xmu.edu.cn; zhouqf@xmu.edu.cn).}
\thanks{Qing Wang is with the School of Computing and Data Science, The University of Hong Kong, Hong Kong, China (email: wangqing@ieee.org).}

% \thanks{\textsuperscript{1}Department of Automation, Xiamen University, Xiamen, China}
% \thanks{\textsuperscript{2}Xiamen Key Laboratory of Big Data Intelligent Analysis and Decision-making, Xiamen University, Xiamen, China}
% \thanks{\textsuperscript{3}School of Engineering, Westlake University, Hangzhou, China}
% \thanks{\textsuperscript{4}Department of Cognitive Service Foundations, IBM T.J. Watson Research Center, New York, NY, USA}

\thanks{\textsuperscript{*}These authors contributed equally to this research.}}
% \thanks{\textsuperscript{†}Corresponding author (e-mail: zhouqf@xmu.edu.cn).}}

% \markboth{This work has been submitted to the IEEE for possible publication.\newline
% Copyright may be transferred without notice, after which this version may no longer be accessible.}%
\markboth{\parbox[t]{20cm}{This work has been submitted to the IEEE for possible publication.\newline Copyright may be transferred without notice, after which this version may no longer be accessible.}}
{Xiao, \MakeLowercase{\textit{et al.}}: 
SS4Rec: Continuous-Time Sequential Recommendation with State Space Models}


% \author[label1,label3]{Wei Xiao\fnref{fn1}}
% \ead{xiaowei@westlake.edu.cn}
% \author[label1,label2]{Huiying Wang\fnref{fn1}}
% \ead{huiyingwang@stu.xmu.edu.cn}
% \author[label1,label2]{Qifeng Zhou\corref{cor1}}
% \ead{zhouqf@xmu.edu.cn}
% \author[label4]{Qing Wang}
% \ead{qing.wang1@ibm.com}
% \affiliation[label1]{%
%   organization={Department of Automation, Xiamen University},
%   city={Xiamen},
%   country={China},
% }
% \affiliation[label2]{%
%   organization={Xiamen Key Laboratory of Big Data Intelligent Analysis and Decision-making, Xiamen University},
%   % \city{Dublin},
%   city={Xiamen},
%   country={China},
% }
% \affiliation[label3]{%
%   organization={School of Engineering, Westlake University},
%   city={Hangzhou},
%   country={China},
% }
% \affiliation[label4]{%
%   organization={Department of Cognitive Service Foundations, IBM T.J. Watson Research Center},
%   city={New York},
%   country={USA},
% }

% \cortext[cor1]{Corresponding author}
% \fntext[fn1]{Both authors contributed equally to this research.}

\maketitle

\begin{abstract}
  Sequential recommendation is a key area in the field of recommender systems aiming to model user interest based on historical interaction sequences with irregular intervals. While previous recurrent neural network-based and attention-based approaches have achieved significant results, they have limitations in capturing system continuity due to the discrete characteristics. In the context of continuous-time modeling, state space model (SSM) offers a potential solution, as it can effectively capture the dynamic evolution of user interest over time. However, existing SSM-based approaches ignore the impact of irregular time intervals within historical user interactions, making it difficult to model complexed user-item transitions in sequences. To address this issue, we propose a hybrid SSM-based model called SS4Rec for continuous-time sequential recommendation. SS4Rec integrates a time-aware SSM to handle irregular time intervals and a relation-aware SSM to model contextual dependencies, enabling it to infer user interest from both temporal and sequential perspectives. In the training process, the time-aware SSM and the relation-aware SSM are discretized by variable stepsizes according to user interaction time intervals and input data, respectively. This helps capture the continuous dependency from irregular time intervals and provides time-specific personalized recommendations. Experimental studies on five benchmark datasets demonstrate the superiority and effectiveness of SS4Rec.
\end{abstract}


\section{Introduction}

Recommender systems are widely employed in social media and e-commerce to alleviate information overload~\cite{wu2022survey, xiao2025pt4rec}. The core of recommender systems is accurately predicting user interests in order to suggest appropriate items for users from a large candidate pool~\cite{wang2019sequential}. Notably, user interests are always changing continuously over time, and this evolution is reflected in their historical interaction sequence. 
As a result, sequential recommender systems have been developed to offer personalized online services by capturing the users' evolving interests through their interaction sequences.

Earlier sequential recommendation approaches utilized Markov chains to capture 
sequential patterns by modeling the transition probabilities between items~\cite{rendle2010factorizing,he2016fusing}. With the advent of deep neural networks, a series of related methods emerged, including those based on recurrent neural networks (RNNs)~\cite{Medsker,hidasi2015session,tan2016improved}, convolutional neural networks (CNNs)~\cite{tang2018personalized}, and Transformer models~\cite{vaswani2017attention,kang2018self,sun2019bert4rec}. Although these approaches achieve promising performance, they are unable to effectively capture long-range dependencies and suffer from the inference inefficiency problem. Additionally, these models are inherently discrete and designed for dealing with discrete systems, however, user interests evolve continuously and vary over time, which requires recommender systems to adapt to the user's current state as well as the dynamic interest evolution.

Recently, state space models (SSMs) have become an alternative to neural network-based models due to their effectiveness in modeling continuous dynamic systems. S4~\cite{gu2021efficiently} and S5~\cite{smith2022simplified} utilize HiPPO matrix initialization and parallel scanning for long-range reasoning and fast inferring, respectively. The latest SSM, Mamba (i.e., S6)~\cite{gu2023mamba}, employs an input-dependent selective state space model to retain input-relevant information while filtering out irrelevant information. This improves its sequential modeling capability and inspires researchers to explore the potential of SSMs for sequential recommendation. Mamba4Rec~\cite{liu2024mamba4rec} and RecMamba~\cite{yang2024uncovering} leverages Mamba to capture sequential dependencies from historical interactions, modeling complex relationships between users and items while partially addressing the trade-off between recommendation performance and inference efficiency.

Despite the success of SSMs in efficient sequential recommendations for long sequences and large-scale datasets, these methods are still under the assumption of uniform time intervals, as Mamba was originally designed for NLP tasks. Therefore, existing SSM-based approaches focus on capturing relation-aware information but overlook the impact of varying temporal dynamics in user interest evolution. In reality, the observed interactions are often recorded as discrete time series data with irregular time intervals~\cite{chen2024contiformer}, which play a crucial role in maintaining the underlying continuity. Although some methods~\cite{li2020time,ye2020time,zhang2023time,wu2020deja} introduce time-dependent embeddings or kernel functions to model temporal information, they still struggle to overcome the assumption of uniform time intervals. This limitation disrupts the underlying continuity of the recommender system, impairing its ability to capture users' evolving interests and time-specific preferences (e.g., morning vs. evening).

To address these challenges, we observe that different SSM architectures possess distinct strengths and propose a hybrid SSM-based model, SS4Rec, for continuous-time sequential recommendation. SS4Rec comprises a \textit{time-aware SSM} and a \textit{relation-aware SSM}, each designed to capture user interests from different perspectives. The discretization steps of these two SSMs are based on the time interval sequences and the user interaction sequences, respectively. Specifically, SS4Rec first employs the time-aware SSM as an effective encoder to handle variable observation intervals, capturing time patterns. It then integrates a selective mechanism into the relation-aware SSM to extract important and relevant information, enhancing its ability to model sequential dependencies. Additionally, residual connections and layer normalization are incorporated to prevent network degradation and ensure training stability.

The major contributions of this work are as follows:
\begin{itemize}
\item We formulate the sequential recommendation problem as a continuous time-varying system and address it using state space models with time-varying parameters. This work eliminates the assumption of uniform time intervals in prior work, allowing us to directly model user interest evolution with irregular-interval interactions, leading to more accurate and time-specific recommendations.
\item SS4Rec is the first work to integrate distinct parameterized state space models to simultaneously capture the continuity of recommender systems from both temporal and sequential perspectives. This design leverages the strengths of SSMs in modeling long-range dependencies while ensuring high inference efficiency.
\item Extensive experiments are conducted on five real-world datasets, and the results show the superiority and effectiveness of SS4Rec compared to other state-of-the-art models.
\end{itemize}

\section{Related Work}

\subsection{Sequential Recommendation}
Sequential recommendation has been extensively studied, with early approaches primarily relying on Markov chain. For example, FPMC (Factorizing Personalized Markov Chains)~\cite{rendle2010factorizing} combines the factorization machine and Markov chain techniques to construct a personalized item-item relationship transition matrix for each user, enabling the next-item prediction based on users' most recent interactions.

In recent years, deep neural networks (DNNs) have gained significant attention due to their strong expressive capabilities, leading to the emergence of DNN-based models in the field of sequential recommendation. GRU4Rec~\cite{hidasi2015session} utilizes Gate Recurrent Unit (GRU)~\cite{Cho} network with ranking loss to capture the user interests within the sequence. NARM~\cite{li2017neural} applies GRU to extract the users' general interests and intentions through a global and local encoder, respectively. SASRec~\cite{kang2018self} employs the Transformer-based architecture to model user interaction sequences, capturing long-term dependencies while adaptively adjusting attention weights of items in the sequence, thereby focusing on important information. Bert4Rec~\cite{sun2019bert4rec} introduces a Cloze task and utilizes a bidirectional attention mechanism to learn contextual information from the interaction sequences. DCRec~\cite{Yang2023DCRec} and GCL4SR~\cite{zhang2022GCL4SR} leverage self-supervised learning techniques to capture complex relationships within sequential data and enhance recommendation performance. All the aforementioned approaches are devoted to modeling sequential dependencies in historical user behaviors but ignore the interaction timestamps, limiting their ability to capture the dynamic evolution of user interests.

\subsection{Time-Aware Recommendation}
% Temporal information is a crucial factor in accurately predicting user interactions. 
Several time-aware sequential recommendation approaches have been proposed to incorporate temporal information in modeling user preferences. TiSASRec~\cite{li2020time} adopts a time-aware self-attentive model to learn the weights of different items in a sequence, integrating both absolute positions and relative time intervals between items to predict future items. TGSRec~\cite{TGSRec} combines sequential patterns and temporal collaborative signals by introducing a Temporal Collaborative Transformer (TCT) layer, capturing both user-item interactions and temporal dynamics for continuous-time recommendation. TAT4SRec~\cite{zhang2023time} processes item sequences and timestamp sequences separately, using a window function to maintain continuity in timestamps. TASER~\cite{ye2020time} introduces a relative temporal attention network to capture the pairwise dependencies using absolute temporal vectors. The work of TiCoSeRec~\cite{dang2023uniform} demonstrates that uniformly distributed time intervals can improve performance. Different from these time-discretized approaches, we explore the continuous-time recommender systems, aiming to extract dynamic evolving information underlying continuous systems.
% CTA, based on self-attention mechanism, manages several parameterized kernels to model the temporal information influence and employ it with contextual information for final determination. Qin et al. present GDERec that designs a graph ordinary differential equation based method to capture the collaborative signals and model the temporal evolution of them. 

\subsection{State Space Models}
State space models have emerged as a promising class of architectures, demonstrating exceptional performance in sequence modeling tasks~\cite{xu2024integrating}. The first structured state space sequence model, S4~\cite{gu2021efficiently}, addresses the challenge of capturing long-range dependencies through the state matrix initialization based on HiPPO framework~\cite{gu2020hippo}. Different from the single-input, single-output scheme of S4, S5~\cite{smith2022simplified} adopts multi-input, multi-output (MIMO) SSM. By utilizing a diagonal state matrix, S5 achieves linear computational complexity and time-varying parameterization via parallel scans. The latest work of SSM, Mamba~\cite{gu2023mamba}, incorporates a selection mechanism to filter irrelevant information while retaining relevant information, and designs a hardware-aware algorithm for both efficient training and inference. Mamba2~\cite{dao2024transformers} connects attention variants with selective SSM directly for large-scale language modeling. Jamba~\cite{lieber2024jamba} is a hybrid architecture for large-scale modeling that combines Mamba with Transformer and mixture-of-experts module to take full advantage of each architecture.

Recent works have introduced Mamba into recommender systems due to its impressive performance in sequential tasks. Mamba4Rec~\cite{liu2024mamba4rec} is the first to apply the Mamba block for sequential recommendation with improving performance and fast inference. RecMamba~\cite{yang2024uncovering} replaces the Transformer block in SASRec with Mamba for lifelong sequences. MaTrRec~\cite{zhang2024matrrec} combines Mamba with Transformer, leveraging their strengths of capturing both long-range and short-range dependencies. SSD4Rec~\cite{qu2025ssd4rec} introduces a bi-directional structured state-space duality block to capture contextual information, similar to Bert4Rec.
Despite these advancements, the challenge of time-specific personalized recommendations with irregular intervals remains largely unexplored. By formulating this problem as a continuous time-varying system and adopting hybrid state space models to capture both temporal and sequential information of user interactions, our work sheds light on the continuous-time sequential recommendation.

\section{SS4Rec}
\subsection{Framework Overview}
In recommender systems, the dynamic changes of user interests can be treated as a continuous time-varying system, where user-item interaction sequences depict the system from discrete states with irregular time intervals.
Our proposed SS4Rec simultaneously models the relationship between user-item interactions (i.e., taking as input points of system) and captures the dynamic evolution of the continuous-time system by incorporating time-aware and relation-aware state space models (SSMs).

\begin{figure*}
    \centering
    \includegraphics[width=0.9\linewidth]{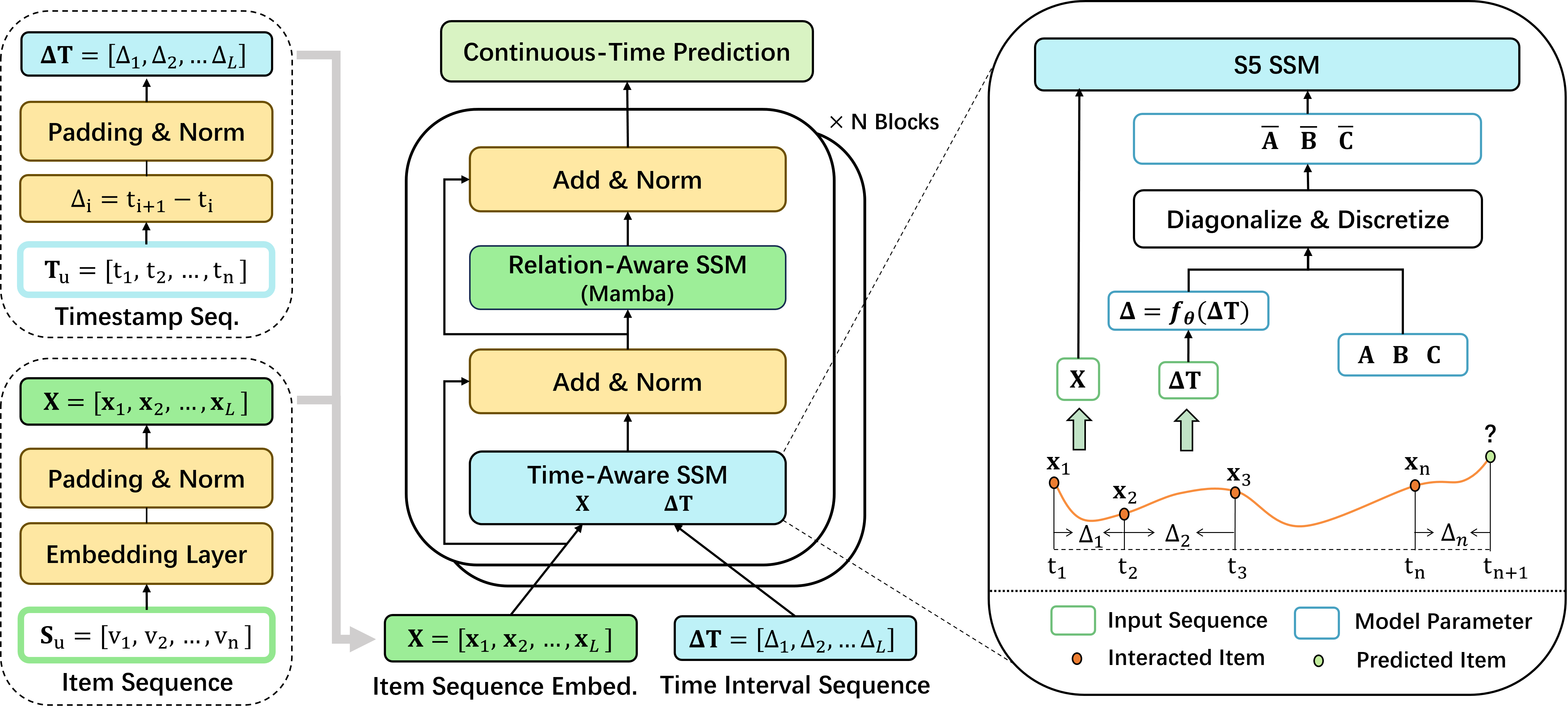}
    \caption{Overview of the SS4Rec framework.}
    \label{fig:framework}
\end{figure*}

An overview of the SS4Rec framework is shown in Fig.~\ref{fig:framework}. SS4Rec is a hybrid model that combines two distinct types of SSMs for sequential recommendation. It consists of initialization, SS4Rec blocks, and continuous-time prediction. Specifically, each SS4Rec block comprises two key components: (1) \textbf{Time-Aware SSM} discretized by variable time steps; (2) \textbf{Relation-Aware SSM} featuring a selection mechanism. Additionally, residual connection and layer normalization are applied between each layer. In the following sections, we will provide a detailed description of the model architecture.

\subsection{Initialization}
In the recommender system, $\mathbf{S}_u=[v_1, v_2, \dots, v_n]$ represents the item sequence of user $u$ and $\mathbf{T}_u=[t_1, t_2, \dots, t_n]$ represents the associated time sequence of user $u$, where $v_i$ denotes the ID of $i$-th item and $t_i$ denotes the interaction timestamp of $v_i$.

At the beginning, the user interaction information is initialized from two aspects. First, the item sequence initialization involves mapping the sequence of items into $D$-dimensional embedding vectors using an embedding layer. In this layer, a learnable item embedding matrix $\mathbf{E} \in \mathbb{R}^{N \times D}$ is employed to project item IDs into high-dimensional vectors, where $N$ refers to the total number of items and $D$ denotes the dimension of the item embedding vectors. The embedding layer then transforms each sequence $\mathbf{S}_u$ into item embedding vectors $\mathbf{X}_u \in \mathbb{R}^{n \times D}$. 
The second aspect is the time interval sequence initialization. Given the time sequence $T_u$ of user $u$, we compute the time intervals $\Delta_i=|t_{i+1}-t_i|$ between two consecutive items $v_i$ and $v_{i+1}$. The time interval sequence for user $u$ is then obtained as $\mathbf{\Delta T}_u=[\Delta_1, \Delta_2, \dots, \Delta_n] \in \mathbb{R}^{n}$. 

For efficient computation, we process sequences in batches to generate the output embeddings $\mathbf{X} \in \mathbb{R}^{B \times L \times D}$ and $\mathbf{\Delta T} \in \mathbb{R}^{B \times L}$, where $L$ represents the maximum length of padded sequences and $B$ refers to the batch size.

\subsection{State Space Model}
State space models (SSMs) are classical tools for modeling system dynamics, with representative works S4, S5, and S6. A SSM converts an input function $x(t)$ to an output function $y(t)$ through an implicit latent state $h(t)$ in the following steps:
\begin{equation}
    h'(t)=\mathbf{A}h(t)+\mathbf{B}x(t)
\label{ssm_1}
\end{equation}
\begin{equation}
    y(t)=\mathbf{C}h(t)
\label{ssm_2}
\end{equation}
where $\mathbf{A} \in \mathbb{R}^{P \times P}$, $\mathbf{B} \in \mathbb{R}^{P \times D}$, $\mathbf{C} \in \mathbb{R}^{D \times P}$ are learnable parameter matrices.
To employ it on a discrete input interaction sequence $[\mathbf{x}_1, \mathbf{x}_2, \dots, \mathbf{x}_n]$ instead of the continuous function $x(t)$, continuous-time SSM in (\ref{ssm_1}) and (\ref{ssm_2}) can be discretized by a constant step size $\mathbf{\Delta}$ through the zero-order hold rule:
\begin{equation}
    \mathbf{h}_t=\bar{\mathbf{A}}\mathbf{h}_{t-1}+\bar{\mathbf{B}}\mathbf{x}_t
\label{ssm_3}
\end{equation}
\begin{equation}
    \mathbf{y}_t=\bar{\mathbf{C}}\mathbf{h}_t
\label{ssm_4}
\end{equation}
where $\bar{\mathbf{A}} \in \mathbb{R}^{P \times P}$, $\bar{\mathbf{B}} \in \mathbb{R}^{P \times D}$, $\bar{\mathbf{C}} \in \mathbb{R}^{D \times P}$ are discretized parameter matrices.

The following two types of SSM represent two ways of converting parameters from the continuous form $(\mathbf{A}, \mathbf{B}, \mathbf{C})$ to the discrete form $(\bar{\mathbf{A}}, \bar{\mathbf{B}}, \bar{\mathbf{C}},\mathbf{\Delta})$, respectively.

\subsection{Time-Aware SSM}
We introduce a time-aware SSM for explicitly capturing temporal information within interaction sequences. It utilizes variable discretization step sizes to capture the dynamic changes of the sequence. This feature allows our method to model user-item interaction sequences in continuous time.

\subsubsection{Diagonalization}
First, we diagonalize the system by definition $\mathbf{A}=\mathbf{V}\mathbf{\Lambda}\mathbf{V}^{-1}$, to compute the linear recurrence using a parallel scan efficiently. $\mathbf{\Lambda}\in \mathbb{R}^{P \times P}$ corresponds to the diagonal matrix containing the eigenvalues and $\mathbf{V}\in \mathbb{R}^{P \times P}$ denotes the eigenvectors.

\subsubsection{Discretization by Variable Timesteps}
The sequence of user interactions contains various temporal intervals. 
These intervals establish the relative positional relationships between items, which are crucial for identifying evolving patterns of user interests. Primitive SSM implementations typically employ a constant timestep, disregarding the variable observation interval $\mathbf{\Delta T}$. Our proposed time-aware SSM discretizes the SSM with variable step $\mathbf{\Delta}$ to maintain the relative temporal relationship of the original data.
It transforms the \textbf{time-invariant} parameters of SSM in (\ref{ssm_3}) and (\ref{ssm_4}) into \textbf{time-varying} parameters:
\begin{equation}
    \mathbf{h}_t=\bar{\mathbf{A}}_t \mathbf{h}_{t-1}+\bar{\mathbf{B}}_t \mathbf{x}_t
\label{ssm_5}
\end{equation}
\begin{equation}
    \mathbf{y}_t=\bar{\mathbf{C}} \mathbf{h}_t
\label{ssm_6}
\end{equation}
where $\bar{\mathbf{A}}=exp(\mathbf{\Delta}\mathbf{\Lambda}) \in \mathbb{R}^{L \times P \times P}$, 
$\bar{\mathbf{B}}=\mathbf{\Lambda}^{-1}(\bar{\mathbf{A}}-\mathbf{I})\mathbf{V}^{-1}\mathbf{B} \in \mathbb{R}^{L \times P \times D}$, 
$\bar{\mathbf{C}}=\mathbf{C}\mathbf{V}  \in \mathbb{R}^{D \times P} $, 
and $\mathbf{\Delta} \in \mathbb{R}^{L \times P}$.

For each user-item interaction sequence, $\mathbf{\Delta T} \in \mathbb{R}^{L}$ denotes its time intervals between consecutive interactions, and $ \mathbf{s} \in \mathbb{R}^{P}$ is the timescale parameter to capture different timescales of each latent channel in the data. The discretization step calculates the product of timescale parameter and time intervals among data, i.e., 
$\mathbf{\Delta} = \mathbf{\Delta T} \cdot \mathbf{s} \in \mathbb{R}^{L \times P}$.

By following the aforementioned process, the time-aware SSM with time-varying parameters is constructed to capture the system dynamics.
% S5 shows excellent performance in long sequence modeling tasks, for its MIMO structure to mix features and its ability to handle irregular time intervals. Moreover, S5 has not been used for recommendation tasks. Therefore, we choose S5 to implement the time-aware SSM in practice. 
In practice, we choose S5 to implement the time-aware SSM due to its proven effectiveness in time sequence modeling tasks. By integrating system dynamics from a temporal perspective, the time-aware SSM layer offers optimized latent representations. This layer facilitates the relation-aware SSM layer in modeling complex and time-varying relationships between items.

\subsection{Relation-Aware SSM}
Selective state space model, Mamba (i.e., S6), is a powerful tool for relation-based modeling, as the replacement of attention mechanism. S6 is a type of state space model with time-varying parameters. In contrast to the aforementioned time-aware SSM in which the timestep $\mathbf{\Delta}$ is dependent on the irregular time intervals $\mathbf{\Delta T}$,
the timestep $\mathbf{\Delta}$ of selective SSM is contingent upon the input data $\mathbf{x}$.

\textit{Selection Mechanism:}
% \subsubsection{Selection Mechanism}
% Compared to previous state space models, S6 layer differs essentially in its time-varying modeling instead of prior time-invariant modeling. 
To model the evolving relationship between the data points, Mamba lets the SSM parameters be functions of the input.
This feature is the ''selection mechanism'' modifying the original SSM to S6. 
Specifically, its parameters $\mathbf{B}$, $\mathbf{C}$, and $\mathbf{\Delta}$ are data-dependent and are transformed by different linear projections of given inputs $ \mathbf{x} \in \mathbb{R}^{B \times L \times D}$:
\begin{equation}
\begin{aligned}
    \mathbf{\Delta} = Softplus(\mathbf{Parameter} + \mathbf{s}_{\mathbf{\Delta}}(\mathbf{x}))& \in \mathbb{R}^{B \times L \times D}, \\
    \mathbf{s}_{\mathbf{\Delta}}(\mathbf{x}) = Broadcast(Linear(\mathbf{x})),
\end{aligned}
\end{equation}
\begin{equation}
    \mathbf{B} = \mathbf{s}_{\mathbf{B}} (\mathbf{x}) = Linear(\mathbf{x}) \in \mathbb{R}^{B \times L \times P},
\end{equation}
\begin{equation}
    \mathbf{C} =\mathbf{s}_{\mathbf{C}} (\mathbf{x}) = Linear(\mathbf{x}) \in \mathbb{R}^{B \times L \times P}.
\end{equation}
$\mathbf{Parameter}$ is a learnable bias summed with $\mathbf{s}_{\mathbf{\Delta}}(\mathbf{x})$.
% Timescale $\mathbf{s}_{\mathbf{\Delta}}$ ensures the selectivity of parameter $\mathbf{A}$. 
The parameters of the selective state space model are discretized by
    $\bar{\mathbf{A}}=exp(\mathbf{\Delta} \mathbf{A}) \in \mathbb{R}^{B \times L \times D \times P}$, 
    $\bar{\mathbf{B}}=(\mathbf{\Delta} \mathbf{A})^{-1}(exp(\mathbf{\Delta} {\mathbf{A})-\mathbf{I})\mathbf{\Delta}\mathbf{B}} \in \mathbb{R}^{B \times L \times P}$, 
    and $\bar{\mathbf{C}}=\mathbf{C} \in \mathbb{R}^{B \times L \times P}$.

The discretized SSM can be obtained by substituting these parameters into (\ref{ssm_5}) and (\ref{ssm_6}).
These parameterizations in the selection mechanism make this SSM layer selectively remember the crucial information and ignore the irrelevant information. 

\subsection{Interpretation of \texorpdfstring{$\mathbf{\Delta}$}{}}
In general, SSM can be interpreted as a continuous system discretized by a timestep $\mathbf{\Delta}$.
The primary distinction of SSMs employed in our framework is the construction of timestep $\mathbf{\Delta}$.
In the case of the original SSM, the timestep is a constant determined by the system sampling frequency. 
For time-aware SSM dealing with time sequences, $\mathbf{\Delta}$ is conditioned on the irregular time interval $\mathbf{\Delta T}$. 
Conversely, for relation-aware SSM, $\mathbf{\Delta}$ is conditioned on the input data $\mathbf{X}$.

\begin{equation}
\left\{
\begin{aligned}
        &\text{Time-Aware SSM: } 
        &\mathbf{\Delta} \leftarrow f_\theta(\mathbf{\Delta T}) \\
        &\text{Relation-Aware SSM: } 
        &\mathbf{\Delta} \leftarrow f_\theta(\mathbf{X}) \\
\end{aligned}
\right.
\end{equation}

Compared to the decoupled two SSMs in this work, another direction worth exploring is the design of a unified SSM whose $\mathbf{\Delta}$ is conditioned on both input data and irregular time interval, i.e., $\mathbf{\Delta} \leftarrow f_\theta(\mathbf{X}, \mathbf{\Delta T})$.

\subsection{Add \& Norm} The residual connection and layer normalization are utilized between each SSM layer for training stability.
The output of each SSM layer is formulated as follows:
\begin{equation}
    \mathbf{X}_{l+1}=LayerNorm(SSM(\mathbf{X}_l)+\mathbf{X}_l)
\label{s5}
\end{equation}
where $\mathbf{X} \in \mathbb{R}^{B \times L \times D}$, $l$ refers to the $l$-th updated layer, and $LayerNorm(\cdot)$ denotes the layer normalization funtion.

\subsection{Continuous-Time Prediction}
Most existing sequential recommenders are next-item predictors, including those that utilize temporal information. This constraint arises from the inherent design of RNNs and Transformers as sequence models optimized for next-token prediction. As a result, RNN-based or Transformer-based sequential recommenders are limited to next-item prediction. They neglect the specific timing of the next interaction, which is crucial for capturing users' time-dependent preferences. Unlike the next-item prediction, our recommender implements continuous-time prediction by the continuous-time parameterization of SSM.

After the stacked SS4Rec blocks, the predictor of our SS4Rec calculates the inner product as a recommendation score based on the next time:
\begin{equation}
    P(v_{next\_time}=v_i)=\mathbf{x_i}^T\mathbf{o}
\label{prediction}
\end{equation}
where $v_{next\_time}$ is the interaction item at time $next\_time$, $\mathbf{o} \in \mathbb{R}^{D}$ denotes the last item embedding in the output embedding of the last SS4Rec block, and $\mathbf{x_i} \in \mathbb{R}^{D}$ denotes the candidate item embedding. In our model, the learned representation contains the information of the next interaction time, i.e., the time when an item needs to be recommended for the user. Thereby, SS4Rec is capable of selecting the most appropriate items at the optimal time for personalized recommendations.

To optimize the SS4Rec model, we adopt the cross entropy as the loss function:
\begin{equation}
    \mathcal{L}=-\sum_{u \in \mathcal{U}} \sum_{i=1}^{N} {y}_{i} log(P(v_{next\_time}=v_i))
\end{equation}
where ${y}_{i}$ refers to the ground truth label of item $i$.

\section{Experiments}
In this section, we present our experimental setup and empirical results in detail.

\begin{table*}[htbp]
\small
\centering
\caption{Datasets Statistics.}
\resizebox{\textwidth}{!}{%
\setlength{\tabcolsep}{3.5pt}
    \begin{tabular}{l|ccccc}
    \toprule
        Dataset & Amazon Sports & Amazon Video-Games & Movielens-1M & KuaiRec(Small) & KuaiRec(Large) \\
        \midrule
        The number of users & 331,845 & 55,145 & 6,040 & 384 & 2,001 \\
        The number of items & 103,912 & 17,287 & 3,952 & 3,328 & 8,975 \\
        The number of inters & 2,835,125 & 496,904 & 1,000,209 & 1,269,384 & 3,497,023 \\
        Average actions of users & 8.54 & 9.01 & 165.60 & 3305.69 & 1748.51 \\
        Sparsity & 99.9918\% & 99.9977\% & 95.81\% & 0.67\% & 80.53\% \\
    \bottomrule
    \end{tabular}%
}
\label{tab:dataset}
\end{table*}

\subsection{Experiments Setting}

\subsubsection{Baselines}
A diverse set of representative sequential recommendation methods is selected for comparison with our method. Reflecting the evolutionary trajectory of sequence model development, these methods have progressed from RNN-based to attention-based mechanisms, with recent advancements exploring SSMs. They are categorized into three groups as follows:

\noindent (1) RNN-based methods
\begin{itemize}
\item \textbf{GRU4Rec}~\cite{hidasi2015session}: it employs Gated Recurrent Unit (GRU) to model user sequences and encode user interests into a final hidden state for session-based recommendation, where each user's sequence of clicks is treated as a session.

\item \textbf{NARM}~\cite{li2017neural}: a GRU-based method adopts an encoder-decoder architecture to capture both user's sequential actions and primary intent within a session.
\end{itemize}
(2) Attention-based methods
\begin{itemize}
\item \textbf{SASRec}~\cite{kang2018self}: this is a classical sequential recommendation method that leverages the Transformer's self-attention mechanism to process the input sequence. 
\item \textbf{BERT4Rec}~\cite{sun2019bert4rec}: it utilizes bidirectional self-attention to model user behavior sequences, allowing each item to incorporated contextual information from both preceding and subsequent interactions in the user’s history.
\item \textbf{TiSASRec}~\cite{li2020time}: this approach introduces discrete time interval embedding in SASRec, enabling it to effectively capture temporal information.
\end{itemize}
(3) SSM-based methods
\begin{itemize}
\item \textbf{Mamba4Rec}~\cite{liu2024mamba4rec}: the first Mamba-based sequential recommender, which mainly replaces the self-attention mechanism with the Mamba layer, in contrast to SASRec.
\item \textbf{RecMamba}~\cite{yang2024uncovering}: this is a simplified model that uses only the mamba layer, which can be referred to as S6-only model in the Ablation Study.
\end{itemize}

\subsubsection{Datasets}
To effectively evaluate the performance of each method, we select five datasets with diverse characteristics, differing significantly in terms of domain, sequence length, time span, and sparsity.

\begin{itemize}
\item \textbf{Amazon}~\cite{mcauley2015image}: a series of datasets are gathered from the Amazon e-commerce platform, comprising a substantial volume of product reviews extracted from the Amazon website. We extract the categories "video games" and "sports" for subsequent experiments.

\item \textbf{Movielens}~\cite{harper2015movielens}: a widely used dataset collected from the Movielens website records user ratings of movies seen. The version (MovieLens-1M) used in our experiments contains 1 million user ratings.

\item \textbf{KuaiRec}~\cite{gao2022kuairec}: a real-world dataset is collected from the recommendation logs of the video-sharing mobile app Kuaishou.
It contains two interaction matrices, a large matrix and a small matrix, where the small matrix, referred to as the first dataset, contains a fully observed user-item interaction matrix.
In this paper, KuaiRec(Large) refers to the large interaction matrix, while KuaiRec(Small) denotes the small one.
\end{itemize}

These datasets span significant periods, with Movielens (1995–2003), Amazon (1996–2018), and KuaiRec (2022) reflecting different stages in the evolution of the Internet. They show considerable variation in both sequence length and sparsity. In terms of average sequence length, the difference between the longest average sequence length (3,305.69 in KuaiRec(Small)) and the shortest average sequence length (8.54 in Amazon Sports) is more than 200 times. Regarding sparsity, Amazon exceeds 99.99\%, while the small matrix of KuaiRec has a sparsity of less than 1\%. Detailed statistics for datasets are provided in Table~\ref{tab:dataset}.

\subsubsection{Implementation Details}
To assess the efficacy of our sequential recommendation model, we follow the standard evaluation protocol used in this field. Specifically, we apply the leave-one-out strategy for testing within the recbole1.0 code framework~\cite{zhao2021recbole}. 
Hit Ratio@K(HR@K), Normalized Discounted Cumulative Gain@K(NDCG@K), and Mean Reciprocal Rank@K(MRR@K) are utilized to evaluate the recommendation performance, with $K = 10$ for all metrics. Higher values for these metrics signify better ranking accuracy, which indicates a more effective recommendation model. 
The specific hyperparameter settings are depicted in Table~\ref{tab:hyperparameter}, and all experiments are conducted on a single A800 80GB GPU.
\begin{table*}[b]
\small
\centering
\caption{Hyper-parameters of SS4Rec.}
    \setlength{\tabcolsep}{6pt}
    \begin{tabular}{l|ccccc}
    \toprule
        Dataset & \makecell[c]{Amazon\\Sports} & \makecell[c]{Amazon\\Video-Games} & \makecell[c]{Movielens\\-1M} & \makecell[c]{KuaiRec\\(Small)} & \makecell[c]{KuaiRec\\(Large)} \\
        \midrule
        Max item sequence length & 50 & 50 & 200 & 200 & 200 \\
        Batch size & 1024 & 1024 & 1024 & 1024 & 1024 \\
        Learning rate & 0.001 & 0.001 & 0.001 & 0.0005 & 0.0005 \\
        Dropout rate & 0.2 & 0.2 & 0.2 & 0.2 & 0.2 \\
        Number of stacked blocks & 3 & 2 & 3 & 3 & 2 \\
        Embedding size & 64 & 64 & 64 & 64 & 64 \\
        Latent size of SSM & 32 & 32 & 32 & 32 & 32 \\
    \bottomrule
    \end{tabular}
\label{tab:hyperparameter}
\end{table*}

\subsection{Overall Comparison}
\begin{table*}[htbp]
\small
\centering
\setlength{\tabcolsep}{6pt}
\caption{Overall performance comparison. The \textbf{bold} indicates the best performance and the \underline{underlined} indicates the second best performance.}
\resizebox{\textwidth}{!}{%
    \begin{tabular}{ll|cccccccc}
    \toprule
        Dataset & Metric & NARM & GRU4Rec & Bert4Rec & SASRec & TiSASRec & Mamba4Rec & SS4Rec\\
        \midrule
        \multirow{3}{*}{Amazon Sports} 
        & HR@10 & 0.0944 & 0.0930 & 0.0846 & 0.1007 & 0.0984 & \underline{0.1028} & \textbf{0.1042} \\
        & NDCG@10 & 0.0826 & 0.0816 & 0.0733 & 0.0839 & 0.0814 & \underline{0.0875} &  \textbf{0.0880} \\
        & MRR@10 & 0.0791 & 0.0781 & 0.0697 & 0.0787 & 0.0761 & \underline{0.0828} & \textbf{0.0830} \\
        \midrule
        \multirow{3}{*}{Amazon Video-Games} 
        & HR@10 & 0.1228 & 0.1145 & 0.1011 & 0.1338 & 0.1238 & \underline{0.1341} & \textbf{0.1362} \\
        & NDCG@10 & 0.0815 & 0.0762 & 0.0672 & 0.0816 & 0.0811 & \textbf{0.0848} & \underline{0.0838} \\
        & MRR@10 & \underline{0.0689} & 0.0646 & 0.0569 & 0.0657 & 0.0680 & \textbf{0.0698} & 0.0678 \\
        \midrule
        \multirow{3}{*}{Movielens-1M} 
        & HR@10 & 0.2636 & 0.2960 & 0.2952 & 0.2505 & 0.2858 & \underline{0.3232} & \textbf{0.3561} \\
        & NDCG@10 & 0.1441 & 0.1694 & 0.1626 & 0.1331 & 0.1587 & \underline{0.1878} &  \textbf{0.2127} \\
        & MRR@10 & 0.1078 & 0.1307 & 0.1224 & 0.0975 & 0.1201 & \underline{0.1466} & \textbf{0.1688} \\
        \midrule
        \multirow{3}{*}{KuaiRec(Small)}
        & HR@10 & 0.2063 & 0.2298 & \underline{0.2324} & 0.2193 & 0.2219 & 0.2141 & \textbf{0.2350} \\
        & NDCG@10 & 0.1444 & 0.1520 & \textbf{0.1587} & 0.1417 & 0.1284 & 0.1456 & \underline{0.1550} \\
        & MRR@10 & 0.1249 & 0.1277 & \textbf{0.1360} & 0.1175 & 0.0989 & 0.1243 & \underline{0.1302} \\
        \midrule
        \multirow{3}{*}{KuaiRec(Large)}
        & HR@10 & 0.1315 & 0.1305 & 0.1315 & 0.1280 & 0.1295 & \underline{0.1415} & \textbf{0.1565} \\
        & NDCG@10 & 0.0788 & 0.0781 & 0.0767 & 0.0755 & 0.0733 & \underline{0.0861} & \textbf{0.0993} \\
        & MRR@10 & 0.0627 & 0.0623 & 0.0600 & 0.0594 & 0.0560 & \underline{0.0691} & \textbf{0.0819} \\
    \bottomrule
    \end{tabular}%
}
\label{tab:performance}
\end{table*}

Our proposed SS4Rec model outperforms most baseline models across datasets with diverse characteristics. Among these baselines, both attention-based and RNN-based methods show comparable performance; Mamba4Rec performs better than the others, except on the KuaiRec(Small) dataset. This indicates that state space models are more effective at modeling sequences and learning user preferences. Furthermore, our model outperforms Mamba4Rec, emphasizing the importance of considering time intervals in sequence recommendations.

In scenarios with varying data densities, our model exhibits different levels of improvement. SS4Rec shows more significant gains in dense datasets, such as Movielens-1M and KuaiRec(Large), compared to sparser datasets like Amazon Sports and Amazon Video-Games. This is because, in denser scenarios, the model can extract a robust data representation from richer information, enhancing its ability to identify underlying patterns. 
In KuaiRec(Small), Bert4Rec achieves optimal performance on NDCG@10 and MRR@10, a result attributed to its bi-directional attention mechanism capturing a greater abundance of supervised learning signals in such a highly dense dataset. 

In summary, our method, SS4Rec, outperforms all baseline approaches across all five datasets when evaluated using the HR@10 metric. Additionally, it achieves the highest performance on three datasets—Sports, Movielens-1M, and KuaiRec(Large)—based on metrics NDCG@10 and MRR@10. For the KuaiRec(Small) and Video-Game datasets, it almost achieves the second-best performance for metrics NDCG@10 and MRR@10. These experimental results demonstrate the effectiveness of SS4Rec in sequential recommendations.

\subsection{Time-Dependent Prediction}
To clearly illustrate SS4Rec's ability for continuous-time prediction, we construct a Toy dataset wherein user interactions with items are exclusively time-dependent. For instance, a user may typically listen to the morning news at 6 a.m. At this specific time, the recommendation system should be able to identify this time-dependent preference and suggest relevant content to the user. 
The construction of the Toy dataset is illustrated as follows in Algorithm~\ref{pseudo-code}.
\begin{algorithm}[htbp]
\caption{Create Toy Dataset}\label{pseudo-code}
\begin{algorithmic}[1]
\State $itemSet \gets [0, 1, \ldots, 99]$
\State $userSet \gets [0, 1, \ldots, 99]$
\State $SeqLen \gets 100$
\State $data \gets []$
\ForAll {$u \in userSet$}
    \State $t \gets \text{RandomInteger}(0, 100)$
    \While {$i < SeqLen$}
        \State $timestamp \gets \text{RandomInteger}(0, 10000)$
        \State $item \gets timestamp\%t$
        \State $data.\text{append}((u, item, timestamp))$
        \State $i = i + 1$
    \EndWhile
\EndFor
\State \Return{$data$}
\end{algorithmic}
\end{algorithm}

Unlike existing discrete-time recommender systems that encode time information through discrete time embeddings, SS4Rec is a continuous-time parametric recommender system. In SS4Rec, time information is encoded through state space functions, allowing the system to capture the dynamic changes of state over time. This continuous-time approach offers higher temporal resolution and flexibility, enabling a more precise understanding of changes in user behavior and preferences.
%In this way, the SS system can more accurately simulate and predict the dynamic behavior patterns of users in continuous time series, thereby providing more personalized recommendations.
To highlight the superiority of our method among time-aware methods, we introduce temporal information into the comparison baselines. Specifically, we introduce a time interval embedding, which is a $D$-dimensional embedding obtained by a single-layer perceptron based on the value of the input time interval. This embedding is then combined with the item embedding.

\begin{figure}[htbp]
    \centering
    \includegraphics[width=1.0\linewidth]{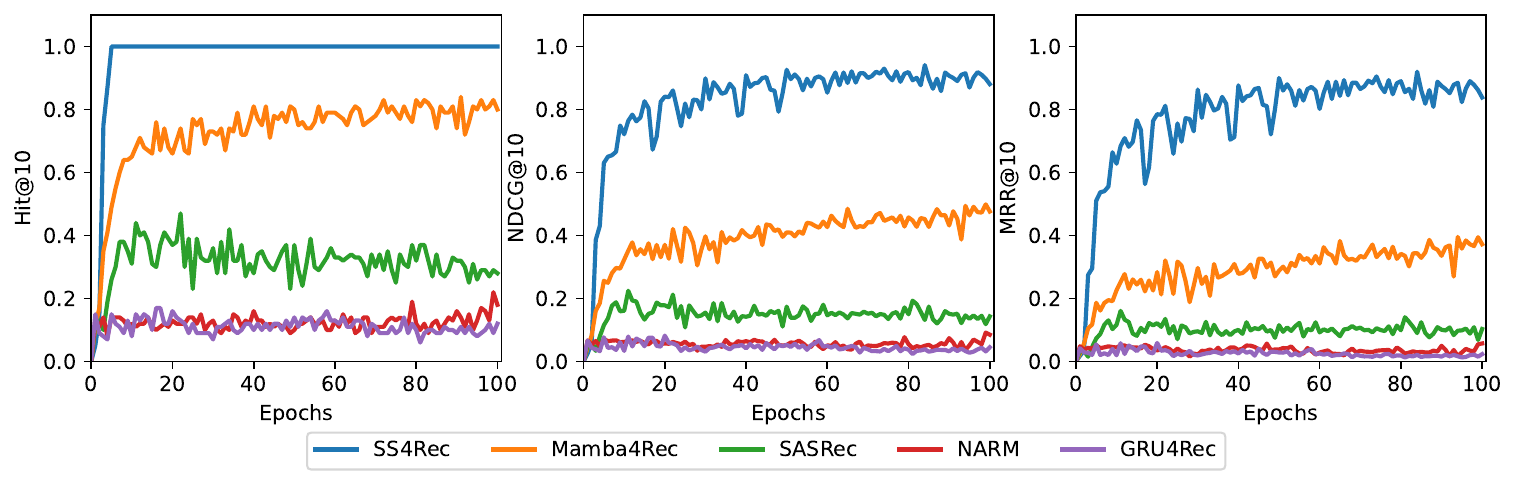}
    \caption{Experiments on Toy dataset. Time-interval embedding is added to item embedding for baselines.}
    \label{fig:toy}
\end{figure}

The results shown in Fig.~\ref{fig:toy} indicate that SS4Rec outperforms methods that utilize discrete-time embeddings. The use of discrete-time embeddings proves insufficient for capturing the dynamics of user interests. In contrast, SS4Rec effectively models the continuously evolving relationships between input points with irregular time intervals in sequences, thus improving time-dependent recommendations.

\subsection{Simulation of Partial Observation}
\begin{figure}[htbp]
    \centering
    \includegraphics[width=1.0\linewidth]{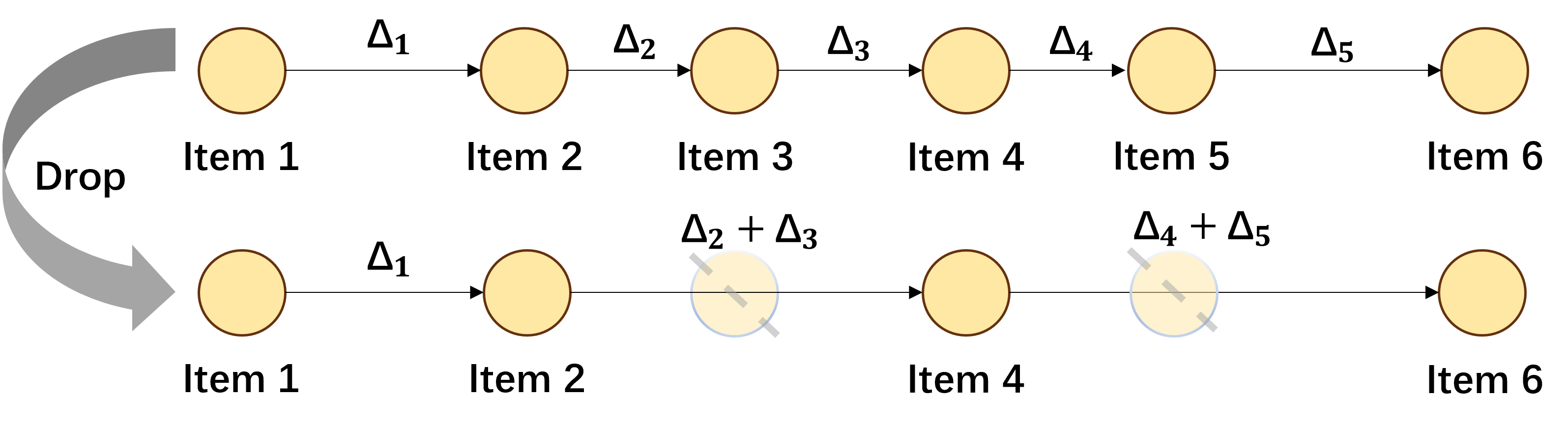}
    \caption{Operation of dropping some items in sequence.}
    \label{fig:drop}
\end{figure}
In real-world scenarios, users often interact with multiple platforms simultaneously. Consequently, the interaction sequence collected may only be a partial observation of the user's complete interaction history. 
To simulate this situation, the items in each sequence are randomly dropped with a 10\% probability in training, as shown in Fig.~\ref{fig:drop}. 
We conduct experiments on randomly dropped sequences to assess the performance of sequential recommendation methods in partially observable sequences. The results are shown in Table~\ref{tab:drop}. 

Compared to RNN-based and attention-based baselines, SSM-based methods consistently outperform them by a considerable margin on most datasets. This demonstrates the effectiveness of using SSM in sequential recommenders. 
Our proposed model, SS4Rec, achieves the best or near-best results demonstrating its adaptability to real-world recommendation scenarios. Even with some observations missing, the model is still capable of utilizing updated time intervals to capture the time variation in sequence. For instance, if \textit{Item 3} is not observed, the time interval between \textit{Item 2} and \textit{Item 4} is updated to $\Delta_2+\Delta_3$. As a result, SS4Rec is better equipped to model the evolution of user preferences over continuous time, enhancing recommendation performance despite partial observations.
Additionally, it is noted that Mamba4Rec performs better on short-sequence (such as Sports and Video-Games), while SS4Rec excels on sequences with longer average lengths and richer temporal information.

\begin{table*}[htbp]
\small
\centering
\setlength{\tabcolsep}{6pt}
\caption{Performance of partially observed sequences. We randomly drop 10\% items for each sequence.}
%\resizebox{\textwidth}{!}{%
    \begin{tabular}{ll|ccccc}
    \toprule
         {Dataset} & {Metric} & {GRU4Rec} & {Bert4Rec} & {SASRec} & {Mamba4Rec} & {SS4Rec} \\
        \midrule
        \multirow{3}{*}{Amazon Sports}
        & HR@10 & 0.0875 & 0.0786 & 0.0932 & \textbf{0.1020} & \underline{0.0975} \\
        & NDCG@10  & 0.0752 & 0.0666 & 0.0761 & \textbf{0.0865} & \underline{0.0803} \\
        & MRR@10  & 0.0714 & 0.0628 & 0.0708 & \textbf{0.0818} & \underline{0.0750} \\
        \midrule
        \multirow{3}{*}{Amazon Video-Games}
        & HR@10 & 0.1069 & 0.0894 & \underline{0.1261} & \textbf{0.1305} & 0.1254 \\
        & NDCG@10 & 0.0692 & 0.0558 & 0.0759 & \textbf{0.0798} & \underline{0.0771} \\
        & MRR@10 & 0.0578 & 0.0455 & 0.0606 & \textbf{0.0643} & \underline{0.0623} \\
        \midrule
        \multirow{3}{*}{Movielens-1M}
        & HR@10 & 0.2823 & 0.2647 & 0.2238 & \underline{0.3103} & \textbf{0.3492} \\
        & NDCG@10 & 0.1596 & 0.1430 & 0.1163 & \underline{0.1781} & \textbf{0.2038} \\
        & MRR@10 & 0.1224 & 0.1059 & 0.0836 & \underline{0.1376} & \textbf{0.1592} \\
        \midrule
        \multirow{3}{*}{KuaiRec(Small)}
        & HR@10 & \textbf{0.2328} & 0.1984 & 0.2141 & 0.2219 & \underline{0.2324} \\
        & NDCG@10 & \underline{0.1563} & 0.1377 & 0.1452 & 0.1428 & \textbf{0.1579} \\
        & MRR@10 & \underline{0.1291} & 0.1188 & 0.1234 & 0.1182 & \textbf{0.1345} \\
        \midrule
        \multirow{3}{*}{KuaiRec(Large)}
        & HR@10 & 0.1370 & 0.1355 & 0.1280 & \underline{0.1605} & \textbf{0.1660} \\
        & NDCG@10 & 0.0820 & 0.0807 & 0.0760 &  \underline{0.0956} & \textbf{0.1042} \\
        & MRR@10 & 0.0652 & 0.0640 & 0.0601 & \underline{0.0758} & \textbf{0.0853} \\
    \bottomrule
    \end{tabular}%
%}
\label{tab:drop}
\end{table*}

\subsection{Ablation Study}
\label{Ablation Study}
We conduct a series of ablation studies to examine each component's contributions to the proposed architecture. Table~\ref{tab:ablation} shows the performance of our default method and its five variants. 

\begin{table}[htbp]
    \centering
    \small
    \setlength{\tabcolsep}{6pt}
    \caption{Ablation analysis.}
        \begin{tabular}{ll|ccc}
        \toprule
            Dataset & Type & HR@10 & NDCG@10 & MRR@10 \\
            \midrule
            \multirow{6}{*}{Movielens-1M} 
            & Ignore & 0.3094 & 0.1747 & 0.1337 \\
            & S5-only & 0.1935 & 0.0968 & 0.0674 \\
            & S6-only & 0.3136 & 0.1831 & 0.1433 \\
            & 1-block & 0.3228 & 0.1895 & 0.1487 \\
            & 2-block & \underline{0.3498} & \underline{0.2059} & \underline{0.1617} \\
            & 3-block & \textbf{0.3561} & \textbf{0.2127} & \textbf{0.1688} \\
            \midrule
            \multirow{6}{*}{KuaiRec(Large)} 
            & Ignore & 0.1430 & 0.0853 & 0.0674 \\
            & S5-only & 0.1210 & 0.0722 & 0.0571 \\
            & S6-only & 0.1420 & 0.0877 & 0.0710 \\
            & 1-block & 0.1365 & 0.0856 & 0.0700 \\
            & 2-block & \textbf{0.1565} & \textbf{0.0993} & \textbf{0.0819} \\
            & 3-block & \underline{0.1530} & \underline{0.0971} & \underline{0.0800} \\
        \bottomrule
        \end{tabular}%
    \label{tab:ablation}
\end{table}

The variants and their respective effects are detailed as follows:
\begin{itemize}
\item \textbf{Ignore time input} Without the dependence on time interval sequence $\mathbf{\Delta T}$, the time-aware SSM layer depends only on item embeddings, i.e., \(\mathbf{\Delta} = \mathbf{s} \in \mathbb{R}^{P}\). It reverts to the same neglect of temporal interval modeling as SASRec and others. A notable decline in the performance of SS4Rec illustrates that time input has a great effect on recommendation. Yet, its performance still surpasses most of the baselines in Table~\ref{tab:performance} and is comparable to Mamba4Rec, demonstrating the inherent advantages of the SSM architecture.

\item \textbf{S5-only} Without the relation-aware SSM layer, performance is significantly worse. The results underscore the critical role of Mamba’s selectivity in processing the output sequence of the time-aware SSM layer.

\item \textbf{S6-only} In this case, the model's architecture is identical to RecMamba. Its outcomes surpass the above variants, yet remain inferior to our proposed method. This indicates the advantage of using the time-aware SSM layer as an encoder, which yields superior representations and enhances the overall performance.

\item \textbf{Number of blocks} Stacking additional blocks with residual connections results in superior performance compared to single block configuration. Stacking two or three blocks achieves the best performance. This suggests that the hierarchical SSM structure facilitates learning more intricate item transitions.
\end{itemize}

\subsection{Efficiency Analysis}
Many previous studies have compared the time complexity of various architectures. In the ideal case of training, SSMs are generally considered to have linear time complexity, while attention-based models are often considered to have squared complexity, as shown in Table~\ref{tab:complexity}.

\begin{table}[htbp]
\centering
\small
\setlength{\tabcolsep}{8pt}
\caption{Time  Complexity. \(L\) and \(d\) represent the sequence length and the dimension of hidden layer respectively.}
    \begin{tabular}{l|c}
    \toprule
        Model & Time Comp. \\
        \midrule
        SSM (SS4Rec/Mamba4Rec) & \(O(L \cdot d^2)\) \\
        Transformer (SASRec/Bert4Rec) & \(O(L^2 \cdot d)\) \\
        RNN (GRU4Rec/NARM) & \(O(L \cdot d^2)\) \\
    \bottomrule
    \end{tabular}
\label{tab:complexity}
\end{table}

\begin{figure}[htbp]
    \centering
    \includegraphics[width=1\linewidth]{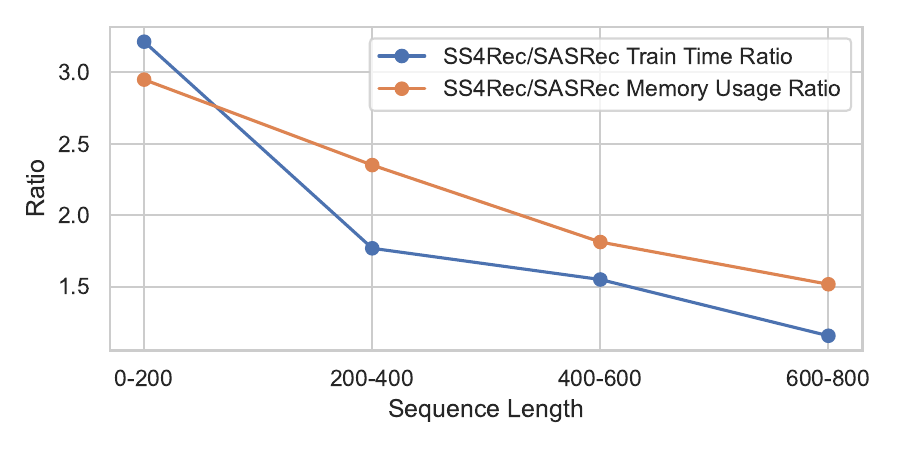}
    \caption{Comparison of practical time/space cost.}
    \label{fig:cost}
\end{figure}

In practical experiments, we found that the Time and Space cost of SS4Rec is larger than that of SASRec. 
In time efficiency, the different discretization steps in the training phase lead to an augmented computational burden.
In terms of spatial complexity, the storage of the timestamps, time intervals, and coefficient matrices after discretization with different step sizes requires much memory. 
A beneficial aspect is that all these costs are linearly related to the sequence length and not squared.
Therefore, to compare the methods' efficiency, we conduct experiments on longer sequence lengths, and Fig.~\ref{fig:cost} reflects the variation of the ratio of SS4Rec/SASRec cost.
As the length of the sequence increases, the difference in computational complexities between linear and squared becomes more apparent. Both temporal and spatial resource consumption of SASRec grows at a faster rate compared to SS4Rec. This implies that the implementation of SS4Rec in recommender systems not only leads to enhanced performance but also exhibits greater efficiency.

\subsection{Performance on Different Lengths of Sequences}
\begin{figure}[htbp]
    \centering
    \includegraphics[width=1\linewidth]{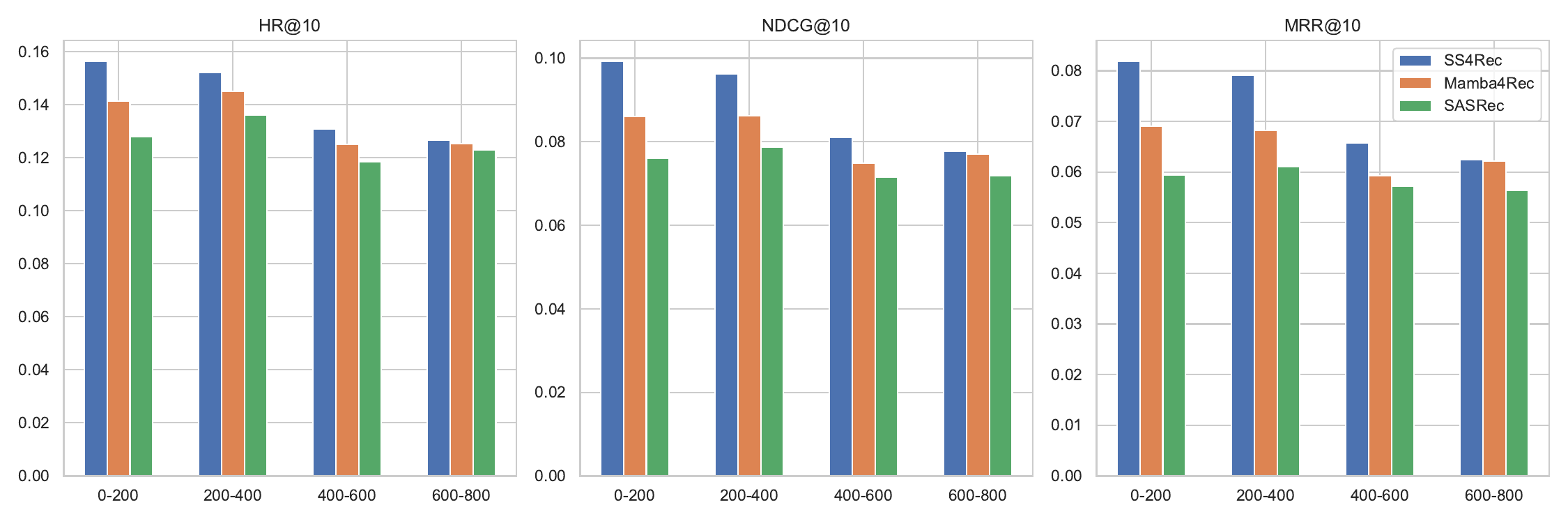}
    \caption{Experiments of different lengths on KuaiRec(Large) dataset.}
    \label{fig:length}
\end{figure}

Fig.~\ref{fig:length} presents the performance of SS4Rec on sequences with different lengths, compared with two representative recommendation models, attention-based model SASRec and SSM-based model Mamba4Rec. The sequences in KuaiRec(Big) dataset are divided into four groups according to their length, including (0,200], (200,400], (400,600], and (600,800], with the amount of data in each group decreasing in order.

We observe that SS4Rec and Mamba4Rec significantly outperform SASRec. This indicates the flexibility of SSM in processing sequences of different lengths. Our proposed SS4Rec model shows optimal performance across all groups. This highlights the importance of integrating temporal information over varying time intervals into SSM, to learn user preferences under continuous time. 
Moreover, comparisons between groups suggest that the performance is affected by the quantity of training data. The more sufficient the data, the higher the performance, and vice versa.

\section{Conclusion and Future Work}
In this paper, we investigate two types of SSMs that employ both time-dependent and input-dependent discretization approaches, to integrate both temporal and sequential information for a more comprehensive understanding of user preferences. Building on this, a hybrid model SS4Rec is proposed for continuous-time sequential recommendation, which effectively models the dynamics of user interest evolution over time. SS4Rec combines the strengths of a time-aware SSM layer for handling variable time intervals and a relation-aware SSM layer for capturing sequential patterns. Additionally, we extend the traditional next-item predictor to a continuous-time predictor, which helps to supply recommendations according to the timing of user interactions. Extensive experiments on five public datasets demonstrate the superiority of our method over state-of-the-art baselines. Furthermore, we conduct several studies to highlight SS4Rec's advantages in continuous-time recommendation and its efficiency in dealing with long interaction sequences.

Our future work will focus on improving the dynamic discretization of time intervals to better capture varying user behavior patterns. We also plan to refine the relation-aware SSM to enhance its ability to prioritize relevant user-item interactions.

%% MOVE TO THANKS
% \section*{Acknowledgements}
% This work was supported in part by the National Natural Science Foundation of China (Grants No. 62171391). Shaorong Fang and Tianfu Wu from Information and Network Center of Xiamen University are
% acknowledged for the help with the GPU computing.
 
% \bibliography{IEEEabrv,./refs}
%%%%%%%%%%%%%%%%%%%%%%%%%%%%%%%%%%%%%%%%%%%%%%%%%%%%%%%%%%%%%%%%%%%%%%%%%%%%%%%%

\bibliographystyle{IEEEtran}
\bibliography{refs}

{\renewcommand*\numberline[1]{Fig.\,#1:\space}
\makeatletter
\renewcommand*\l@figure[2]{\noindent#1\par}
\makeatother

\listoffigures}

\vfill

\end{document}